\documentclass{JAC2003}

\usepackage[dvips]{graphicx}
\usepackage{booktabs}
\usepackage{subfigure}

\begin{document}
\title{LHCb BEAM-GAS IMAGING RESULTS}

\author{P. Hopchev, LAPP, IN2P3-CNRS,\\
Chemin de Bellevue, BP110, F-74941, Annecy-le-Vieux\\
For the LHCb Collaboration}

\maketitle


\begin{abstract}
The high resolution of the LHCb vertex detector makes it possible to perform
precise measurements of vertices of beam-gas and beam-beam interactions and
allows beam parameters such as positions, angles and widths to be determined.
Using the directly measured beam properties the novel beam-gas imaging method
is applied in LHCb for absolute luminosity determination. In this contribution
we briefly describe the method and the preliminary results obtained with May
2010 data.
\end{abstract}

\section{INTRODUCTION}
The methods for measuring the absolute luminosity are generally divided into
indirect and direct ones. Some of the indirect methods to be used at the LHC are:
\begin{Itemize}
    \item Optical Theorem - can be used to determine the absolute luminosity
          without knowing the beam intensities. ALTAS-ALFA~\cite{cite-alfa} and 
          TOTEM~\cite{cite-totem} are going to measure the elastic scattering of protons
          with detectors located about 200 m away from IP 1 and 5 respectively.
    \item Use precisely calculable process. For example in e+/e- colliders the 
          Bhabha scattering process is used. At LHC the most promising candidates are the 
          QED processes of Z and elastic muon pair production, both of which are going
          to be used in LHCb. A more complete description of the prospects for these
          measurements can be found elsewhere in these proceedings~\cite{cite-janderson}.
    \item Reference cross-section - once any absolute cross-section has been measured
          it can be used as a reference to calculate the cross-section for other processes.
\end{Itemize}
    
The direct methods determine the luminosity by measuring the beam parameters:
\begin{Itemize}
    \item Wire method~\cite{cite-wire-method} - scan thin wires across the beams and measure rates.
    \item Van der Meer method~\cite{cite-vdm-method} - scan beams across each other and measure rates.
    \item Beam-gas imaging method~\cite{cite-massi-bgmethod} - reconstruct beam-gas interaction
          vertices to measure the beam angles, positions and shapes. Results from the application
          of this method are discussed in this contribution.
    \item Beam imaging during van der Meer scan - a recently proposed method~\cite{cite-vladik-VDMnote}
          to measure the beam profiles and overlap by vertex reconstruction of beam-beam interactions.
\end{Itemize}

For absolute luminosity normalization in 2010 the LHC and each of its large experiments
performed van der Meer scans. Detailed descriptions of the procedure and the results can be
found elsewhere in these proceedings~\cite{cite-vdm-alice, cite-vdm-atlas, cite-vdm-cms, cite-vdm-lhcb}.

The beam-gas method for absolute luminosity determination was proposed by 
M. Ferro-Luzzi in 2005 and was applied for a first
time in LHCb (~see \cite{cite-lhcb-kshort, cite-vladik-moriond, cite-plamen-moriond}~) using
the first LHC data collected in the end of 2009. This measurement represents the only absolute
cross-section normalization performed by the LHC experiments at 900 GeV.

In this contribution we report on the further application and results of the method, using 7 TeV
center-of-mass energy data collected by LHCb in 2010. Apart from the common beam current measurement,
the beam-gas imaging provides an absolute luminosity normalization, which is
independent from the one obtained with the van der Meer method.

\section{BEAM-GAS IMAGING METHOD}
The luminosity for a single pair of counter-rotating bunches can be expressed with the
following general formula~\cite{cite-Napoly}:
\begin{equation}\label{ref_eq_lumi_formula}
L ~ = ~ f N_{1} N_{2} K \int{\rho_{1}(\vec{r},t)\rho_{2}(\vec{r},t) ~ d^{3}\vec{r} ~ dt},
\end{equation}
where $f$ is the bunch revolution frequency, $N_{i}$ are the number of particles in the colliding bunches,
\mbox{$K = \sqrt{(\vec{v_1} - \vec{v_2})^2 - \frac{(\vec{v_1} \times \vec{v_2})^2}{c^2}}$} is the M{\o}ller
kinematic relativistic factor, $c$ is the speed of light, $\vec{v_{i}}$ are the 
bunch velocities and $\rho_{i}(\vec{r},t)$ are the bunch densities, normalized such that
their integral over full space is equal to 1 at any moment~t: $\int{\rho_{i}(\vec{r},t) ~ d^{3}\vec{r} ~ dt} = 1$.

As described in~\cite{cite-vladik-VDMnote}, for the case of no crossing angle,
the luminosity formula can be written as a function only of the transverse profiles
of the colliding bunches $\rho_{i}^{\perp}(x,y)$:
\begin{equation}\label{ref_eq_lumi_formula_perp}
L ~ = ~ f N_{1} N_{2} \int{\rho_{1}^{\perp}(x,y)\rho_{2}^{\perp}(x,y) ~dx~dy}
\end{equation}
The effect on the luminosity for the case of non-collinear beams is described later, in 
the section {\it Analysis Overview}. The beam-gas imaging method aims at measuring the overlap
integral for a given bunch-pair by measuring the angles, offsets and transverse
profiles of the two colliding bunches. This is achieved by 
reconstructing beam-gas interaction vertices. The gas used as a {\sl visualizing medium}
can be the residual gas in the beam vacuum pipe, which consists mainly of relatively light atoms like
hydrogen, carbon and oxygen, or a specially designed gas-injection system can be used to
create a controlled pressure bump in the region of the LHCb vertex detector. The later
would allow to perform the beam profile measurements in a shorter time, thus reducing
the effects from potential beam instability. In addition, the injection of gas with high
atomic number, like xenon, will result in high multiplicity interaction vertices and
improved primary vertex resolution.

An important prerequisite for the proper reconstruction of the bunch profiles is the 
transverse homogeneity of the visualizing gas. A dedicated test performed in October 2010
measured the beam-gas interaction rates as function of beam displacement in a plane
perpendicular to the beam axis. The beams were moved within $\pm 150 ~ \mu$m (approximately 
3 times the beam width) from their nominal position in both x and y. This allowed us to set a
limit on the distortion of the measured beam profiles due to transverse inhomogeneity of
the residual gas. The needed beam overlap correction from a non-uniform transverse distribution
of the residual gas was found to be smaller than 0.05\% and was neglected.

The principal precision limitations of the beam-gas method are:
\begin{Itemize}
    \item Vertex resolution - its knowledge plays increasingly important role
          as the beam sizes become smaller than the resolution.
    \item Beam-gas rate - determines the time needed to {\sl snapshot} the beam profiles 
          and the associated statistical uncertainty.
    \item Beam stability - in case of fluctuations of the beam orbits and sizes non-trivial
          systematic effects need to be taken into account.
\end{Itemize}  

It is important to note that in contrast to the van der Meer method the beam-gas imaging 
method does not involve movement of the beams. This means that possible beam-beam effects
are constant and potential effects which depend on the beam displacement, like
hysteresis, can be avoided. Furthermore, the beam-gas imaging method is applicable
during physics fills.

\section{PRELIMINARY RESULTS FROM MAY 2010 DATA}
\begin{figure}[htb]
   \centering
   \includegraphics*[width=82.5mm]{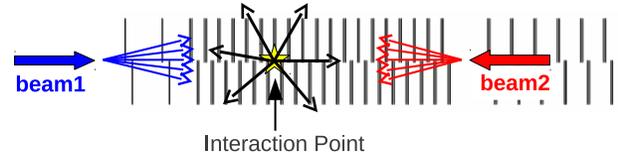}
   \caption{A sketch of the VELO, including the 2 Pile-Up stations on the left. The thick arrows 
            indicate the direction of the LHC beams (beam1 going from left to right), while the 
            thin ones show example directions of flight of the products of the beam-gas and 
            beam-beam interactions.}
   \label{ref-fig-velo}
\end{figure}

LHCb \cite{cite-lhcb} is a forward spectrometer covering the pseudo-rapidity range of $\eta \in [2;5]$.
It is equipped with a vertex detector (Vertex Locator, VELO), positioned around the interaction point.
The VELO consists of two retractable halves, each having 21 modules of radial and azimuthal
silicon-strip sensors with half-circle shape, see Fig.~\ref{ref-fig-velo}. Its excellent acceptance
for beam-gas and beam-beam interactions is determined by its length of almost a meter and 
the small inner radius of the sensors, which approach the beam to merely 8 mm when the VELO
is at its nominal, closed position. The two most upstream stations (left side of figure~\ref{ref-fig-velo}), 
the so called {\it Pile-Up System}, are used in the Level-0 trigger. The VELO is the sub-detector
essential for the application of the beam-gas imaging method at LHCb.

\subsection{Running Conditions and Beam-gas Trigger}
The data used for the results described in this contribution was taken in May 2010
when there were between 2 and 13 bunches per beam and the number of colliding pairs at
LHCb varied between 1 and 8. The trigger included a dedicated selection for events
containing beam-gas interactions. The relevant hardware (Level-0)
triggers are:

\begin{Itemize}
    \item Beam1-gas: select events with a Calorimeter transverse energy sum larger 
          than 3 GeV and a Pile-Up System multiplicity lower than 40. 
    \item Beam2-gas: select events with a Calorimeter transverse energy sum smaller 
          than 6 GeV and a Pile-Up System multiplicity larger than 9. 
\end{Itemize}

These triggers were enabled in all b-e and e-b crossings (throughout this contribution
'e' is used for denoting an empty bunch slot and 'b' - a bunch slot filled with protons).
For the colliding bunches no beam-gas Level-0 trigger was used and the beam-gas events
were selected only if they passed any of the 'physics' trigger channels or if they
happen to coincide with a proton collision, which fired any of the Level-0 trigger channels.
In May 2010 the LHCb hardware trigger was non-selective and the beam-gas interactions in b-b crossings
were triggered efficiently. At the High Level Trigger a simple proto-vertexing algorithm 
selected events by looking for accumulation of tracks around a point on the z axis. The 
same algorithm was used for the b-e, e-b and b-b crossings, but different z-selection 
cuts were applied. For example during b-b crossings only interactions with $z < - 350$~mm or 
$z > 250$~mm were selected.

\subsection{VELO Vertex Resolution}
\begin{figure}[t!]
  \begin{center}
    \vspace{5mm}
    \subfigure[Beam-beam resolution in x and y.]{
      \includegraphics*[width=82.5mm]{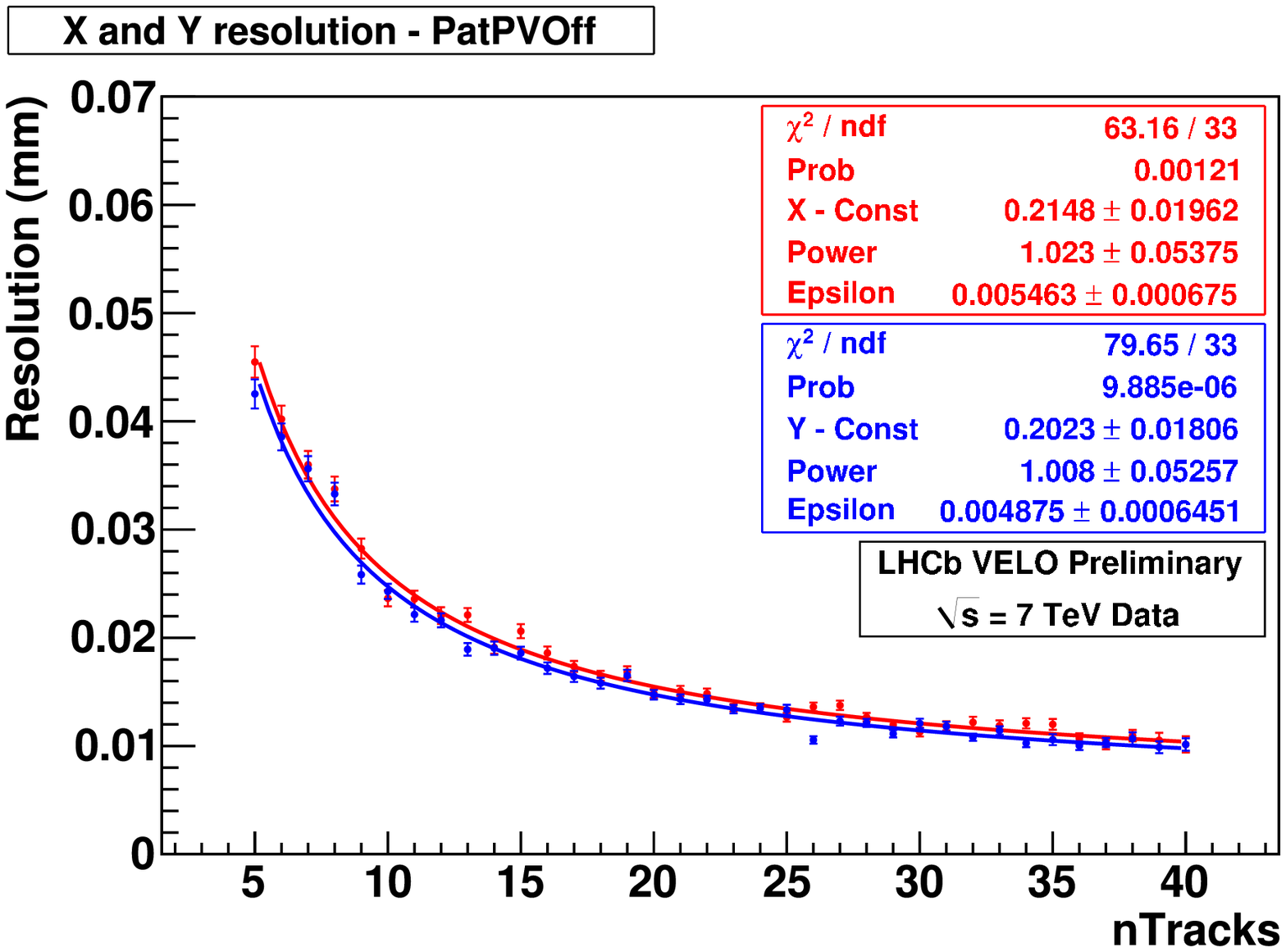}
      \label{ref-fig-veloRes1} }
    \vspace{5mm}
    \subfigure[Beam1-gas resolution correction factor in x and y.]{
      \includegraphics*[width=75.0mm]{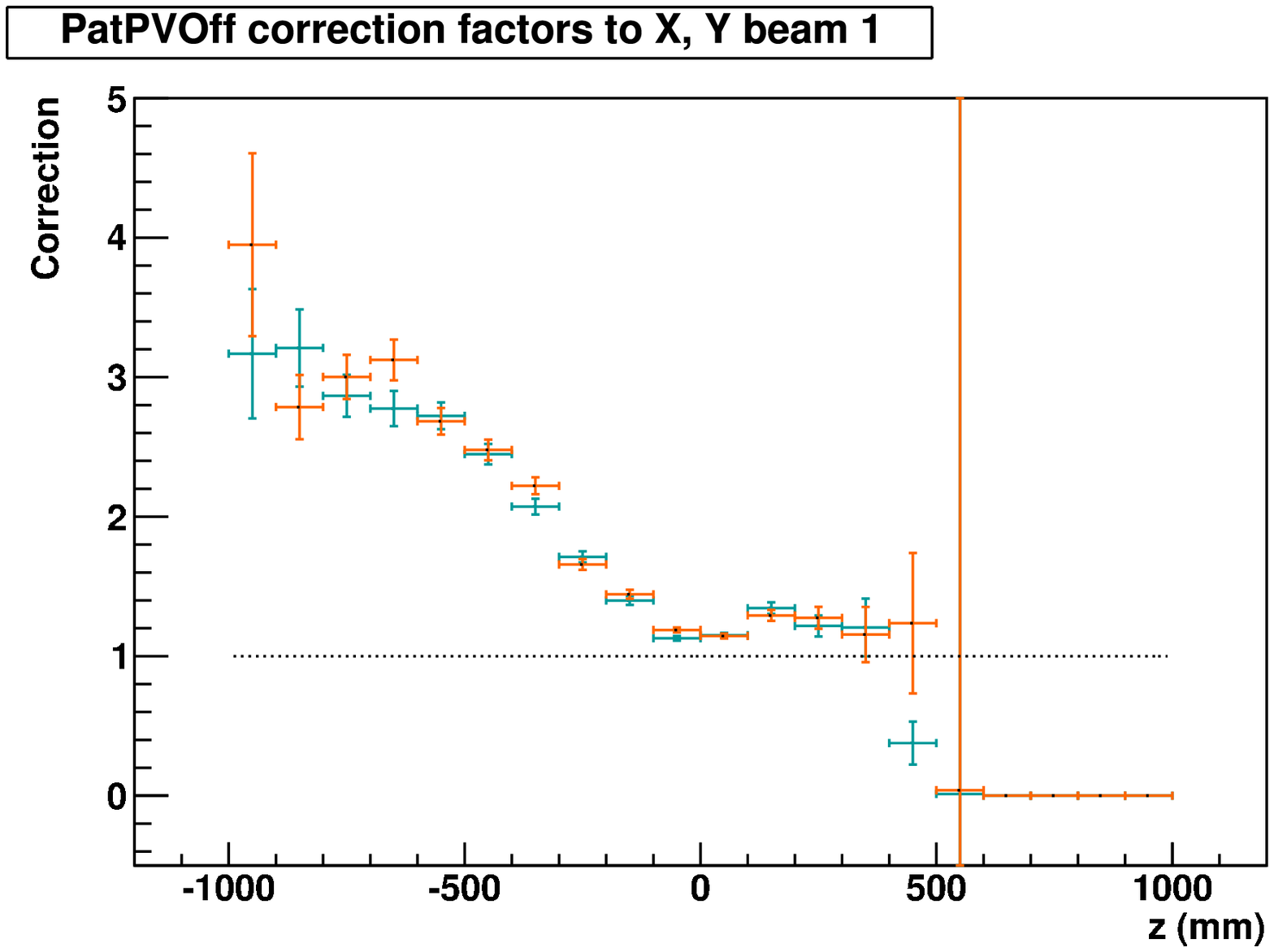}
      \label{ref-fig-veloRes2_1} }
    \subfigure[Beam2-gas resolution correction factor in x and y.]{
      \includegraphics*[width=75.0mm]{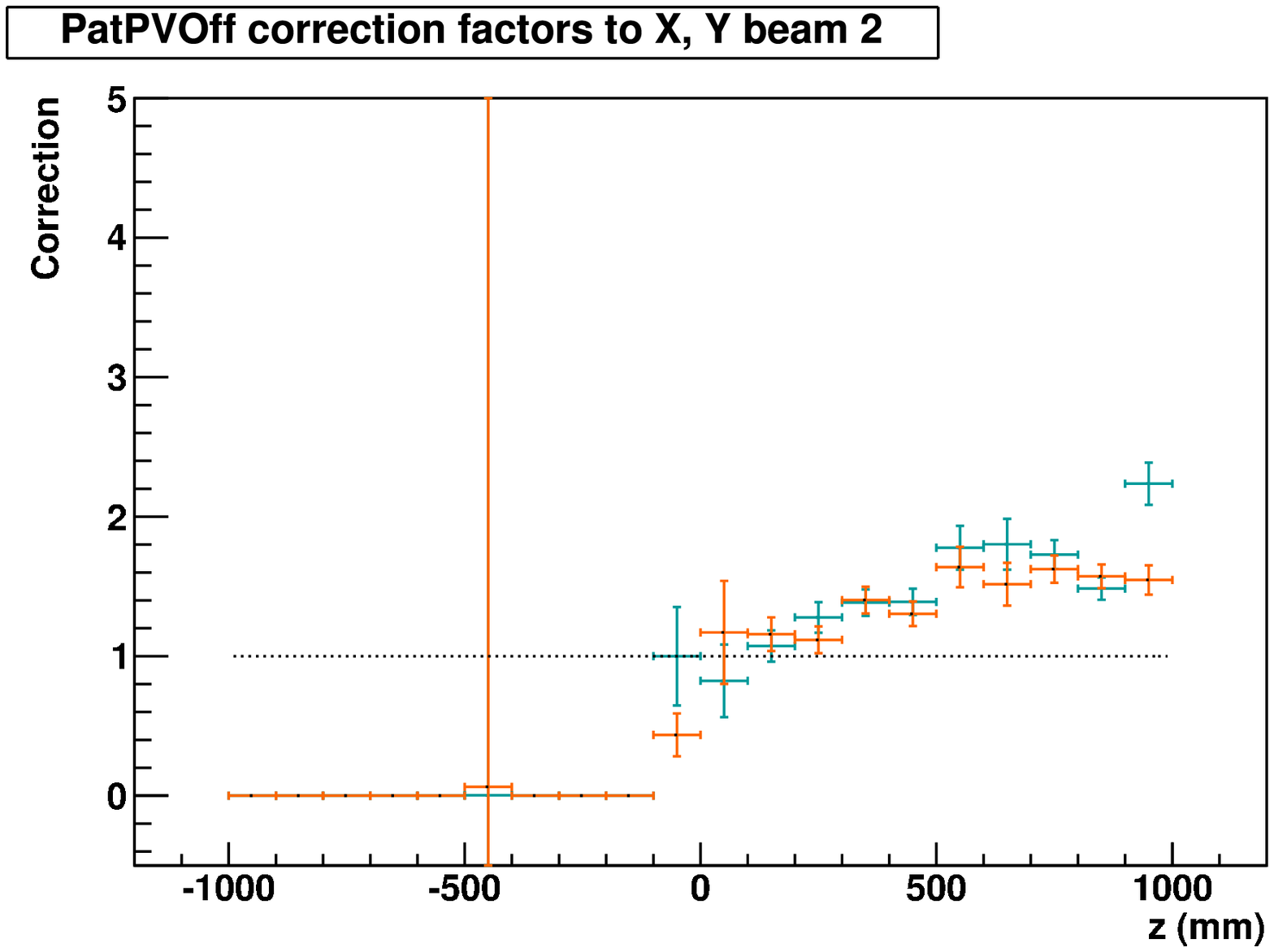}
      \label{ref-fig-veloRes2_2} }
  \end{center}
  \caption{(a) VELO primary vertex resolution in the transverse directions x and y 
    for beam-beam interactions. (b) and (c) resolution corrections in x and y for beam1-gas
    and beam2-gas interactions, accounting for the z-dependence of the resolution.}
\end{figure}

The VELO primary vertex resolution was determined from data in the following way. We randomly split
the reconstructed VELO tracks in two equal samples, run a vertexing algorithm on each of them and
require that the two reconstructed vertices have equal number of tracks. The width of the distribution
of the distance between the two vertices divided by $\sqrt{2}$ gives the resolution estimate for
the half-track vertices. The resolution is parametrized with a Gaussian in two steps. First we
estimate the resolution for beam-beam interactions as function of the number of tracks in the vertex,
see Fig.~\ref{ref-fig-veloRes1}. The used parametrization function has the following form:
\begin{equation}\label{ref_eq_res_param}
R(N) = \frac{\sigma_0}{N^{0.5+\frac{\delta}{N^2}}} + \epsilon,
\end{equation}
where $\sigma_0$ is a parameter determining the resolution for small number of tracks, 
$N$ is the number of tracks per vertex, the power $\delta$ accounts for the deviation
from the $1/\sqrt{N}$ behavior and $\epsilon$ is the asymptotic resolution for large number of tracks
per vertex. Later, by comparing the resolution for beam-gas and beam-beam vertices with the same
number of tracks, we calculate a correction factor which takes into account the z-dependence of the
resolution. Fig.~\ref{ref-fig-veloRes2_1} and Fig.~\ref{ref-fig-veloRes2_2} show the beam-gas
correction factor as function of z. Finally, the parametrized vertex resolution is used to unfold
the true size of the beams.

\subsection{Analysis Overview}
\label{sec-analysis}
\begin{figure}[htb]
   \centering
   \includegraphics*[width=82.5mm]{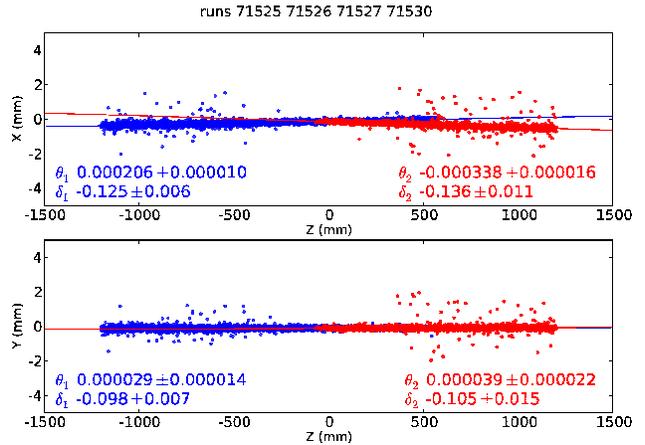}
   \caption{Position of reconstructed beam-gas interaction vertices during \mbox{b-e} and \mbox{e-b} crossings
            for a May 2010 fill. The measured crossing angles in the horizontal and vertical planes
            (544 $\pm$ 26 and -10 $\pm$ 36 $\mu$rad respectively) agree reasonably well with the 
            expectations (540 and 0 $\mu$rad respectively).}
   \label{ref-fig-beamAngles}
\end{figure}

To measure the beam positions and transverse profiles we plot the position of the beam-gas vertices
in the x-z and y-z planes. In Fig.~\ref{ref-fig-beamAngles} we show an example for the case of \mbox{b-e} 
and \mbox{e-b} crossings. The straight line fits provide the beam angles in the corresponding planes.
In general we observe an agreement between the expected and measured beam angles. It is important to 
note that the colliding bunches are the only relevant ones for the luminosity measurement, because we 
need to measure the overlap integral for the colliding bunch-pairs.

\begin{figure}[htb]
  \begin{center}
    \subfigure[Bunches of beam1. Top: x profile, Bottom: y profile.]{ \includegraphics*[width=82.5mm]{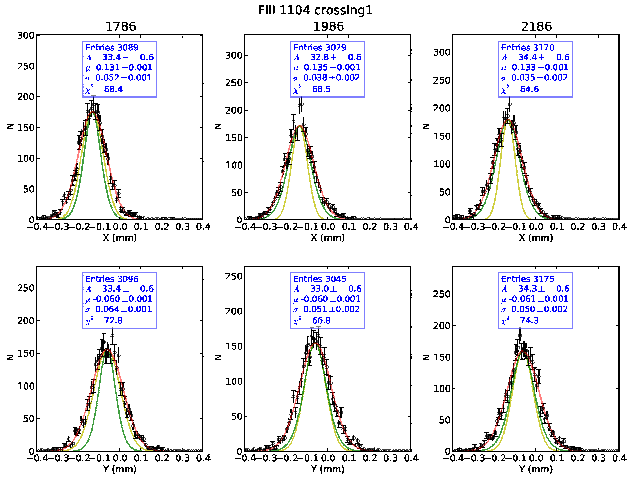} }
    \subfigure[Bunches of beam2. Top: x profile, Bottom: y profile.]{ \includegraphics*[width=82.5mm]{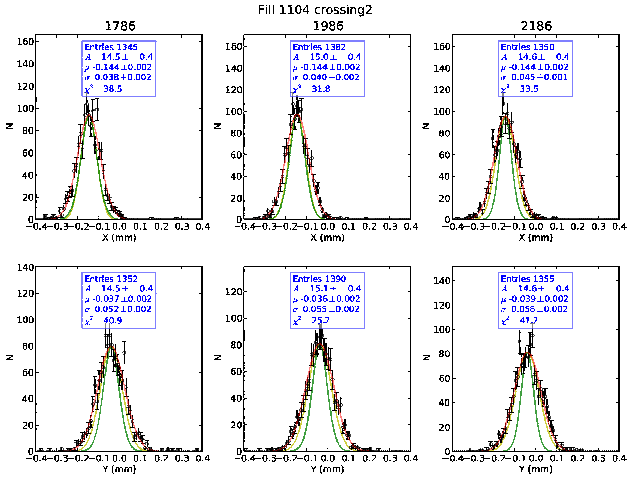} }
  \end{center}
  \caption{x and y profiles of three colliding bunch-pairs. The different lines represent the raw measured
     size, the vertex resolution and the unfolded size.}
   \label{ref-fig-beamProfiles}
\end{figure}

The bunch x and y profiles are obtained from the projection of the x-z and y-z beam-gas vertex
distributions onto a plane perpendicular to the beam direction. As an example
in Fig.~\ref{ref-fig-beamProfiles} we show the x and y profiles of three colliding bunch-pairs.

The true bunch size is obtained after deconvolving the vertex resolution. Equally importantly, 
we use the following relations between the position and size of the luminous region
($\mu_{BB}$ and $\sigma_{BB}$) and the positions and sizes of the individual beams
($\mu_{1}$, $\mu_{2}$, $\sigma_{1}$ and $\sigma_{2}$) as a constraint in the fits.
\begin{equation}
\sigma_{BB}^{2} = \frac{\sigma_1^2 \sigma_2^2}{\sigma_1^2 + \sigma_2^2} \;\;\;\;\;\;\;\;\;
\mu_{BB} = \frac{\mu_1 \sigma_2^2 + \mu_2 \sigma_1^2}{\sigma_1^2 + \sigma_2^2}
\end{equation}
This improves significantly the precision of the beam size measurements.

The calculation of the overlap integral is done initially by integration of the product of 
two Gaussians representing the widths and positions found with the procedure mentioned above.
As the beam profiles are measured in a plane perpendicular to the beam direction we have
not yet taken into account the fact that the bunches are tilted and do not collide head-on. 
The crossing angle correction to be applied on the overlap integral can be approximated with
the following formula:
\begin{equation}\label{ref-eq-angleCor}
C_{\hbox{crossing angle}} = \sqrt{1+\left( \frac{\theta_c \, \sigma_z}{\sigma_x} \right) ^2},
\end{equation}
where $\theta_c$ is the half crossing angle and $\sigma_x$ and $\sigma_z$ are the bunch sizes
in the crossing angle plane. The longitudinal beam size is measured from the beam spot assuming
that the two beams have equal size. For the beam conditions in May 2010 the crossing angle
overlap correction factor was about 0.95. Formula \ref{ref-eq-angleCor} is a good approximation for
the case of no transverse offsets and equal bunch sizes. We now use a numerical calculation
which makes a small difference (~0.5\%). 

\subsection{Preliminary Results with May 2010 Data}
\begin{figure}[htb]
   \centering
   \includegraphics*[width=82.5mm]{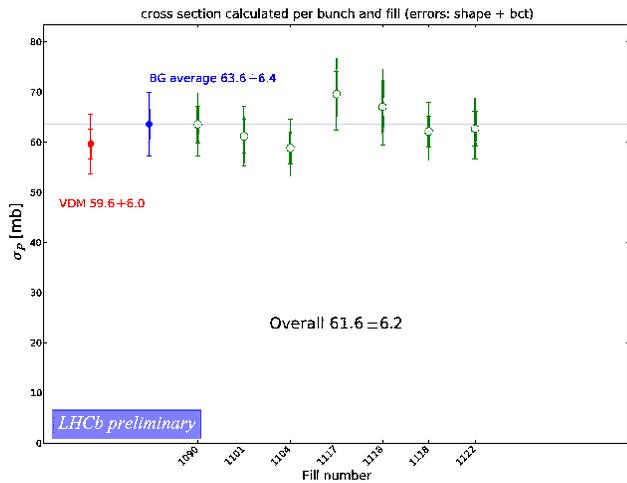}
   \caption{Preliminary results for an LHCb-specific cross-section measured with 
     the beam-gas imaging method. The results for seven different physics fills
     taken in May 2010 are compared with the ones obtained from a van der Meer scan
     performed in April 2010.}
   \label{ref-fig-lumi}
\end{figure}

With the use of the beam-gas imaging method and following the outlined procedure we
performed seven independent measurements of an LHCb-specific reference cross-section.
The beam currents, essentially needed for this method, were obtained following the
same procedure as described in~\cite{cite-bcnwg-note1}. The main uncertainties contributing to
the overall precision of the cross-section measurement come from the bunch widths (3\%),
their relative positions (3\%) and crossing angle (1\%). The measurement of the beam
intensities was done with precision of 5\% and has dominant contribution to the
overall uncertainty. The preliminary results of the analysis are summarized in Fig.~\ref{ref-fig-lumi}.
For multi-bunch fills the results obtained for each colliding pair were averaged.

The absolute scale knowledge is propagated through the full LHCb dataset with the
use of several independent luminosity monitors.

\section{CONCLUSIONS}

The beam-gas imaging method was applied on data collected by LHCb in May 2010
and provided an absolute luminosity normalization with uncertainty of 10\%,
dominated by the knowledge of the beam intensities. The measured LHCb-specific
cross-section is in agreement with the measurement performed with the van der
Meer method.

In the beginning of 2011 a more refined analysis was performed, providing an improved
precision of the beam parameters measured by LHCb. Most notably this later analysis
profited from an improved knowledge of the beam currents which allowed a significant
reduction of the absolute luminosity normalization uncertainty.

Further precision improvements are possible in dedicated fills with broader beams
or fills where both beam-gas and van der Meer methods can be applied simultaneously.
In a addition, a controlled pressure bump in the LHCb interaction region would
allow us to apply the beam-gas imaging method in a shorter time, decreasing the effects
from beam instabilities.


\end{document}